\documentclass[journal]{IEEEtran}

\title{Channel Characteristics for Molecular Communications with an Absorbing Receiver}

\usepackage{amssymb}

\ifCLASSINFOpdf
  \usepackage[pdftex]{graphicx}
\else
  \usepackage[dvips]{graphicx}
\fi
\usepackage[cmex10]{amsmath}
\usepackage[tight,footnotesize]{subfigure}

\usepackage{color}

\newcommand{\disp}{\displaystyle}
\newcommand{\ba}{\begin{array}}
\newcommand{\ea}{\end{array}}
\newcommand{\btab}{\begin{table}}
\newcommand{\etab}{\end{table}}
\newcommand{\bcen}{\begin{center}}
\newcommand{\ecen}{\end{center}}
\newcommand{\btabb}{\begin{tabular}}
\newcommand{\etabb}{\end{tabular}}
\newcommand{\bea}{\begin{eqnarray}}
\newcommand{\eea}{\end{eqnarray}}
\newcommand{\beqn}{\begin{equation}}
\newcommand{\eeqn}{\end{equation}}
\newcommand{\beqnt}{\begin{equation*}}
\newcommand{\eeqnt}{\end{equation*}}
\newcommand{\bex}{\begin{example}}
\newcommand{\eex}{\end{example}}

\newcommand{\rrn}{r_{r}}
\newcommand{\receiver}{\Omega_{r}}

\newcommand{\prt}{p(r,t|r_0)}
\newcommand{\prtat}[1]{p(#1,t|r_0)}
\newcommand{\urt}{u(r,t|r_0)}
\newcommand{\vrt}{v(r,t|r_0)}
\newcommand{\prtzero}{p(r,t \rightarrow 0 |r_0)}

\newcommand{\cdfNhit}{N_{\text{hit}}(\receiver, t|r_0)}
\newcommand{\cdfNhitt}[1]{N_{\text{hit}}(\receiver, #1|r_0)}
\newcommand{\pdfNhit}{n_{\text{hit}}(\receiver, t|r_0)}
\newcommand{\dortPiDe}{\sqrt{4\pi D t^3}}
\newcommand{\expoPiDe}{\,e^{-d^2/4Dt}}

\newcommand{\ntxs}{N_{\text{Tx}}^{r_0}}

\newcommand{\nrxts}[2]{N_{\text{Rx}}^{\receiver}(#1,#2)}
\newcommand{\nrxtsOm}[2]{N_{\text{Rx}}^{\Omega}(#1,#2)}

\newcommand{\pulsept}{t_{\text{peak}}}
\newcommand{\pulsepn}{n_{\text{peak}}}

\begin{document}
%
\title{3-D Channel Characteristics for Molecular Communications with an Absorbing Receiver}
%
%
%

\author{H. Birkan~Yilmaz,
        Akif Cem~Heren,
        Tuna~Tugcu,~\IEEEmembership{Member,~IEEE,}
        and\\~Chan-Byoung~Chae,~\IEEEmembership{Senior Member,~IEEE}
\thanks{H. B. Yilmaz and C.-B. Chae are with the School of Integrated Technology, Yonsei University, Korea. e-mail: \{birkan.yilmaz, cbchae\}@yonsei.ac.kr}
\thanks{A. C. Heren and T. Tugcu are with Bogazici University, Istanbul, Turkey. e-mail: \{akif.heren, tugcu\}@boun.edu.tr}
}

%
%

\markboth{IEEE Communications Letters}%
{3-D Channel Characteristics for Molecular Communications with an Absorbing Receiver}
%



\maketitle

\begin{abstract}
Within the domain of molecular communications, researchers mimic the techniques in nature to come up with alternative communication methods for collaborating nanomachines. This work investigates the channel transfer function for molecular communication via diffusion. In nature, information-carrying molecules are generally absorbed by the target node via receptors. Using the concentration function, without considering the absorption process, as the channel transfer function implicitly assumes that the receiver node does not affect the system. In this letter, we propose a solid analytical formulation and analyze the signal metrics (attenuation and propagation delay) for molecular communication via diffusion channel with an absorbing receiver in a 3-D environment. The proposed model and the formulation match well with the simulations without any normalization.
\end{abstract}


%
\IEEEpeerreviewmaketitle

\section{Introduction}

\IEEEPARstart{N}{anomachines} are defined as the devices ranging in size from 0.1 $\mu m$ to 10 $\mu m$ and composed of nano-scale components in size less than 100 $nm$ at least in one dimension \cite{tarakanov2010carbon}. Cooperation and communication between such machines enable the realization of complex applications such as health monitoring, tissue engineering, nanomedicine, and environment monitoring \cite{akyildiz2011nanonetworks, nakano2012molecular}. According to the IEEE P1906.1 definition, nanonetworking deals with the communication between nano- and/or micro-scale machines that has at least one nano-scale component, and is controlled or engineered by humans. 

In the literature, various molecular communication systems, such as molecular communication via diffusion (MCvD), calcium signaling, microtubules, pheromone signaling, and bacterium-based communication are proposed \cite{akyildiz2011nanonetworks, nakano2012molecular}. Among these systems, MCvD is an effective and energy-efficient method for transporting information \cite{kuran2010energy, kim2013novelModulation}. In MCvD, the information is transmitted via the propagation of molecules through the environment. An MCvD system is composed of five main processes: encoding, emission (transmission), propagation, absorption (reception), and decoding~\cite{Chae_JSAC14}. In nature, most of the receptor types remove the information-carrying molecules from the environment once they arrive at the receiver, which means molecules contribute to the signal once in a short duration. Therefore, considering absorption is necessary since using concentration or the molecule distribution function as the channel transfer function implicitly assumes the receiver node does not affect the MCvD system. 

In this letter, we derive the first hitting probability function in a 3-D environment. To the best of our knowledge, it is the first analytical 3-D channel characterization in the nanonetworking literature for an absorbing receiver. Prior work using the concentration function for the channel characterization requires a normalization factor to agree with the simulation results, which necessitates simulation. In this letter, the first hitting probability distribution for an absorbing spherical receiver is introduced with a solid formulation, which matches well with the simulations without any normalization. Based on our analytical and numerical results, we confirm that an inverse Gaussian distribution with a modification is valid even for a $\mbox{3-D}$ environment. Note that final analytical formula, which will be introduced in Subsection~\ref{subsec:fpp}, is not a distribution since there is a positive probability of not hitting on the absorbing boundary for a diffusing particle in a 3-D environment when time goes to infinity. 


\section{System Model}

\subsection{Molecular Communication via Diffusion}
\begin{figure*}[!t]
\centering
\includegraphics[width=1.99\columnwidth,keepaspectratio] {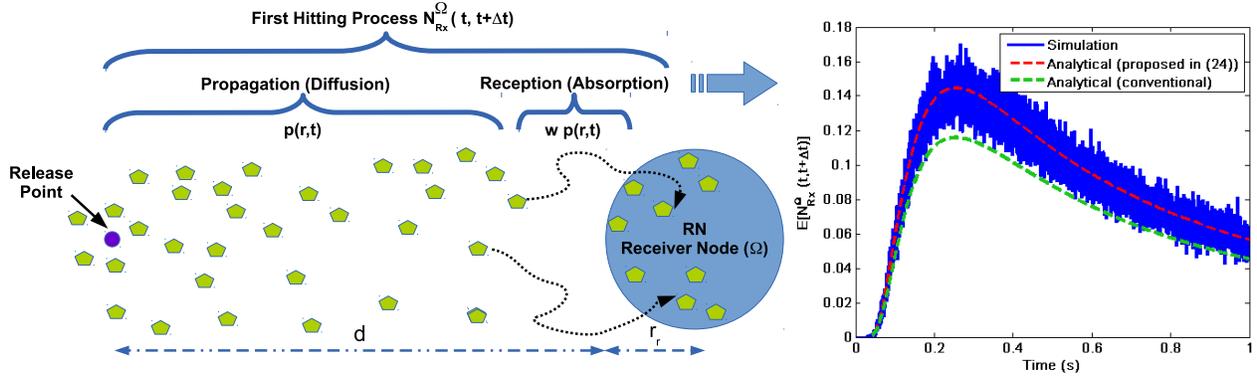}
\caption{ MCvD system and $\nrxtsOm{t}{t+\Delta t}$ plot (${\ntxs = 5000}$ molecules, ${\rrn = 10\mu m}$, ${D = 79.4\mu m^2/s}$, ${d = 10\mu m}$, ${\Delta t = 10^{-4}}$). Evaluations in this letter are carried out using the human insulin hormone like molecules as the information-carrying molecule in a blood like fluid at 310$^{\circ}$K.}
\label{fig:mcvdSystem}
\end{figure*}
The physical system considered in this letter consists of a molecular channel, a transmitter-receiver pair, and information-carrying molecules. Fig. \ref{fig:mcvdSystem} depicts a diagram of the system. The molecular channel is in a 3-D environment, represented by a spherical coordinate space and has infinite extent in all dimensions. This environment is completely filled with a fluid of viscosity $\eta$ and is exempt from flow currents. Therefore, the propagation in the environment solely depends on Brownian motion. The transmitter is a point source of a size equal to zero and is located at distance $r_0$ from the center of the absorbing receiver and distance $d$ from the closest point on the surface of the receiver, where $d = r_0 - \rrn$. Messenger molecules are non-trivial chemical compounds (such as proteins or polypeptides), which are emitted from the point transmitter to carry the encoded information to the receiver. 
After their release, each molecule moves independently within the molecular channel according to Brownian motion. The receiver is a 3-D sphere of radius $\rrn$ with fully absorbing boundaries. That is, every molecule colliding with the surface of the sphere is absorbed by the receiver body and removed from the communication environment. The receiver is assumed to have the ability to count the number of absorbed molecules in any given time interval.

\subsection{Microscopic Theory of Diffusion}
The microscopic theory of diffusion can be developed from two simple assumptions. The first is that a substance will move down its concentration gradient. A steeper gradient results in the movement of more particles. The derivative of the flux with respect to time results in Fick's second law in a 3-D environment
\beqn
\label{eq:ficks3}
\frac{\partial \prt}{\partial t} = D \nabla^2 \prt
\eeqn
\noindent where $\nabla^2$, $\prt$, and $D$ are the Laplacian operator, the molecule distribution function at time $t$ and distance $r$ given the initial distance $r_0$, and the diffusion constant, respectively. The value of $D$ depends on the temperature, viscosity of the fluid, and the Stokes' radius of the molecule \cite{tyrrell1984diffusion}.

\section{Channel Characteristics}

\subsection{Hitting Rate to Spherical Absorber}
\label{subsec:fpp}
For the calculation of the hitting rate to a spherical absorber, we consider the methodology presented in \cite{redner2001guide} and \cite{schulten2000lectures}. In the remainder of this section, we present the derivation of the formulas to ultimately find the fraction of molecules absorbed by the receiver as a function of time.

In addition to Fick's 3-D diffusion equation, we should define the initial and the boundary conditions obeying the problem at hand.

The initial condition is defined as
\begin{equation}
\prtzero = \frac{1}{4 \pi r_0^2} \delta(r - r_0),
\end{equation}
and the first boundary condition is
\begin{equation}\label{eq:boundary_inf}
\lim_{r \rightarrow \infty} \prt = 0,
\end{equation}
which denotes the assumption that the distribution of the molecules vanishes at distances far greater than $r_0$. The second boundary condition is 
\begin{equation}\label{eq:boundary_sphere}
D \frac{\partial \prt}{\partial r} = w \prt \text{ , for } r = \rrn
\end{equation}
for which as $w$, the rate of reaction, approaches infinity we create a boundary where every collision leads to absorption. When we have an absorption for every collision, we consequently have a diminishing $\prt$ as approaching to the surface of the absorber (i.e., $\prtat{\rrn} = 0$, which also holds as a boundary condition).

First we observe that Fick's second law (\ref{eq:ficks3}) becomes 
\begin{equation}\label{eq:radial_diffusion}
\frac{\partial (r \cdot \prt)}{\partial t} = D \frac{\partial^2 r \cdot \prt}{\partial r^2}
\end{equation}
when we move to the spherical coordinates and drop terms with $\theta$ and $\phi$ from the Laplacian operator since $\prt$ is spherically symmetric and depends solely on $r$. 

As the next step, we partition $\prt$ into two equations, $\urt$ and $\vrt$, which individually obey the radial diffusion equation \eqref{eq:radial_diffusion} and together obey the boundary conditions \eqref{eq:boundary_inf} and \eqref{eq:boundary_sphere}. 
Therefore, the function $\urt$ must satisfy
\begin{align}
\frac{\partial (r \cdot \urt)}{\partial t} &= D \frac{\partial^2(r \cdot \urt)}{\partial r^2} \\
\label{eq:fourier_initial_condition} r \cdot u(r, t \rightarrow 0|r_0) &= \frac{1}{4\pi r_0} \delta(r - r_0).
\end{align}
Through Fourier transform, we obtain:
\begin{align} \label{eq:fourier_U}
r \cdot \urt = \frac{1}{2\pi} \int\limits_{-\infty}^{+\infty} \mathcal{U}(k, t|r_0) e^{ikr} dk.
\end{align}
When we plug \eqref{eq:fourier_U} into \eqref{eq:radial_diffusion}, we get:
\begin{align}
\mathcal{U}(k, t|r_0) = K_u \cdot \exp[- D t k^2]
\end{align}
where $K_u$ is the time-independent coefficient and is determined from the initial condition \eqref{eq:fourier_initial_condition} as:

\begin{equation}
K_u = \frac{1}{4\pi r_0} e^{-i k r_0},
\end{equation}
which results in the final Fourier expression:
\begin{align}
r \cdot \urt = \frac{1}{8\pi^2 r_0} \int\limits_{-\infty}^{+\infty} \exp[-Dtk^2] e^{ik(r-r_0)} dk
\end{align}
that yields the following expression after integration:
\begin{equation}
r \cdot \urt = \frac{1}{4 \pi r_0} \frac{1}{\sqrt{4 \pi D t}} \exp \left[ - \frac{(r-r_0)^2}{4Dt} \right].
\end{equation}
Secondly, we handle the remaining part, $\vrt$, which must satisfy
\begin{align}
\frac{\partial (r \cdot \vrt)}{\partial t} &= D \frac{\partial^2(r \cdot \vrt)}{\partial r^2} \\
r \cdot v(r, t \rightarrow 0|r_0) &= 0.
\end{align}
Through Laplace transform this time, we have
\begin{equation}
\frac{s}{D} (r \cdot \mathcal{V}(r, s|r_0)) = \frac{\partial^2(r \cdot \mathcal{V}(r, s|r_0)}{\partial r^2}
\end{equation}
where $\mathcal{V}(r, s|r_0)$ is the Laplace transform of $\vrt$. Applying the boundary condition \eqref{eq:boundary_inf}, we obtain
\begin{equation}
r \cdot \mathcal{V}(r, s|r_0) = K_v \exp \left[ - \sqrt{\frac{s}{D}} r \right]
\end{equation}
where $K_v$ is a constant that should satisfy the second boundary condition \eqref{eq:boundary_sphere}. At this point, we consider the transform of the complete solution $\prt$ rather than the inverse Laplace transform of $\vrt$ since from the boundary condition it is easier to determine the arbitrary constant $K_v$. Here, we finally plug in the Laplace transform of $\urt$ and obtain
\begin{align}
r \cdot \mathcal{P}(r, s|r_0) &= r \cdot \mathcal{U}(r, s|r_0) + r \cdot \mathcal{V}(r, s|r_0), \nonumber\\ 
                              &= \frac{1}{4\pi r_0} \frac{1}{\sqrt{4Ds}} ~e^{ - \sqrt{\frac{s}{D}} |r-r_0|}  
							  K_v ~e^{ - \sqrt{\frac{s}{D}} r }.
\label{eq:general_laplace}
\end{align}
Also with the Laplace transform of the boundary condition~\eqref{eq:boundary_sphere}, we get
\begin{equation}\label{eq:boundary_laplace}
\left. \frac{\partial (r \cdot \mathcal{P}(r, s|r_0))}{\partial r} \right\vert_{r = \rrn} = \frac{w \rrn + D}{D \rrn} \rrn \cdot \mathcal{P}(r, s|r_0).
\end{equation} 
From \eqref{eq:general_laplace} and \eqref{eq:boundary_laplace}, we determine
\begin{align}
K_v &= \frac{\sqrt{s/D} - (w \rrn +D) / (D \rrn)}{\sqrt{s/D} + (w \rrn +D) / (D\rrn)} \frac{1}{4\pi r_0} \frac{1}{\sqrt{4Ds}} \nonumber \\
& \times  \exp \left[ - \sqrt{s/D} (r_0 - 2 \rrn) \right]
\end{align}
which produces the ultimate result for $\mathcal{P}(r, s|r_0)$. The inverse Laplace transform of $\mathcal{P}(r, s|r_0)$ yields to
\begin{align} \label{eq:prt_initial}
\prt &= \frac{1}{4\pi r r_0} \frac{1}{\sqrt{4 \pi D t}} \left( \exp \left[- \frac{(r-r_0)^2}{4Dt}\right] \right. \nonumber \\ 
&+ \left. \exp \left[- \frac{(r + r_0 - 2 \rrn)^2}{4Dt}\right]  \right) \nonumber \\
&-  \frac{1}{4\pi r r_0} \frac{w\rrn + D}{D\rrn} \exp \left[ \left(\frac{w \rrn + D}{D\rrn}\right)^2 Dt \right. \\ 
&+  \left. \frac{w\rrn + D}{D\rrn} (r + r_0 - 2\rrn) \right] \nonumber \\
& \times \text{erfc} \left[ \frac{w\rrn + D}{D\rrn} \sqrt{Dt} + \frac{r + r_0 - 2\rrn}{\sqrt{4Dt}} \right]. \nonumber
\end{align}

Now for the case of the absorbing boundary, we consider the case when $w \rightarrow \infty$ and the solution \eqref{eq:prt_initial} becomes
\begin{equation}
\prt = \frac{1}{4\pi r r_0} \frac{1}{\sqrt{4 \pi D t}}  \left( e^{- \frac{(r-r_0)^2}{4Dt}} - e^{- \frac{(r + r_0 - 2\rrn)^2}{4Dt}} \right).
\end{equation}

Following the molecule distribution $\prt$, we find the hitting rate of the molecules $\pdfNhit$ to the receiver $\receiver$, which is given by
\bea
\label{eqn:pdfNhit}
\pdfNhit &=& 4 \pi \rrn^2 w p(\rrn, t|r_0) \nonumber\\
         &=&  \displaystyle\frac{\rrn}{r_0} \frac{1}{\sqrt{4 \pi D t}} \frac{r_0 - \rrn}{t} ~e^{ - \frac{(r_0 - \rrn)^2}{4Dt} }. 
\eea

Furthermore, by integrating $\pdfNhit$ we obtain $\cdfNhit$, which is the fraction of molecules absorbed by the receiver until time $t$:
\begin{align}
\cdfNhit &= \int\limits_{0}^{t} n_{\text{hit}} (\receiver, t'|r_0)  dt' \nonumber \\
         &= \frac{\rrn}{r_0} \text{erfc} \left[\frac{r_0 - \rrn}{\sqrt{4Dt}}\right]
\label{eqn:cdfNhit}
\end{align}
Hence, the expected number of molecules hitting to the receiver in an interval $[t, t+\Delta t]$ can be evaluated by 
\begin{align}
\begin{split}
\mathbb{E}[\nrxts{t}{t+\Delta t}] &= \ntxs \, \left\{ \cdfNhitt{t+\Delta t} \right. \label{eq:fig} \\
                         &- \left. \cdfNhitt{t} \right\}
\end{split}
\end{align}
where $\mathbb{E}[.]$ and $\ntxs$ denote the expectation operator and the number of emitted molecules, respectively.

The main contribution of this letter is to introduce (\ref{eqn:pdfNhit}) and~(\ref{eqn:cdfNhit}) to the nanonetworking domain. By utilizing $\pdfNhit$ and $\cdfNhit$, we can understand MCvD channel response in a 3-D environment. Therefore, these formulations are crucial for system designers and researchers. We investigate and reveal two important channel metrics in the following subsections.

\subsection{Pulse Peak Time}
MCvD signal has one peak as shown in Fig.~\ref{fig:mcvdSystem}. Hence we can find the mean pulse peak time, $\pulsept$, by finding the vanishing point for the derivative of $\pdfNhit$ with respect to time: 
\beqn
  \frac{\partial \pdfNhit}{\partial t}  = \disp\frac{\partial \left( \frac{\rrn}{r_0} \frac{d}{\dortPiDe} \expoPiDe \right) }{\partial t} = 0.
\label{eqn:pulsePeakTime}
\eeqn
Solving (\ref{eqn:pulsePeakTime}) in terms of $t$ leads us to 
\beqn
\mathbb{E}[\pulsept] = \frac{d^2}{6D}.
\label{eqn:pulsePeakTime2}
\eeqn

\begin{figure}[!t]
\centering
\includegraphics[width=0.9\columnwidth,keepaspectratio] {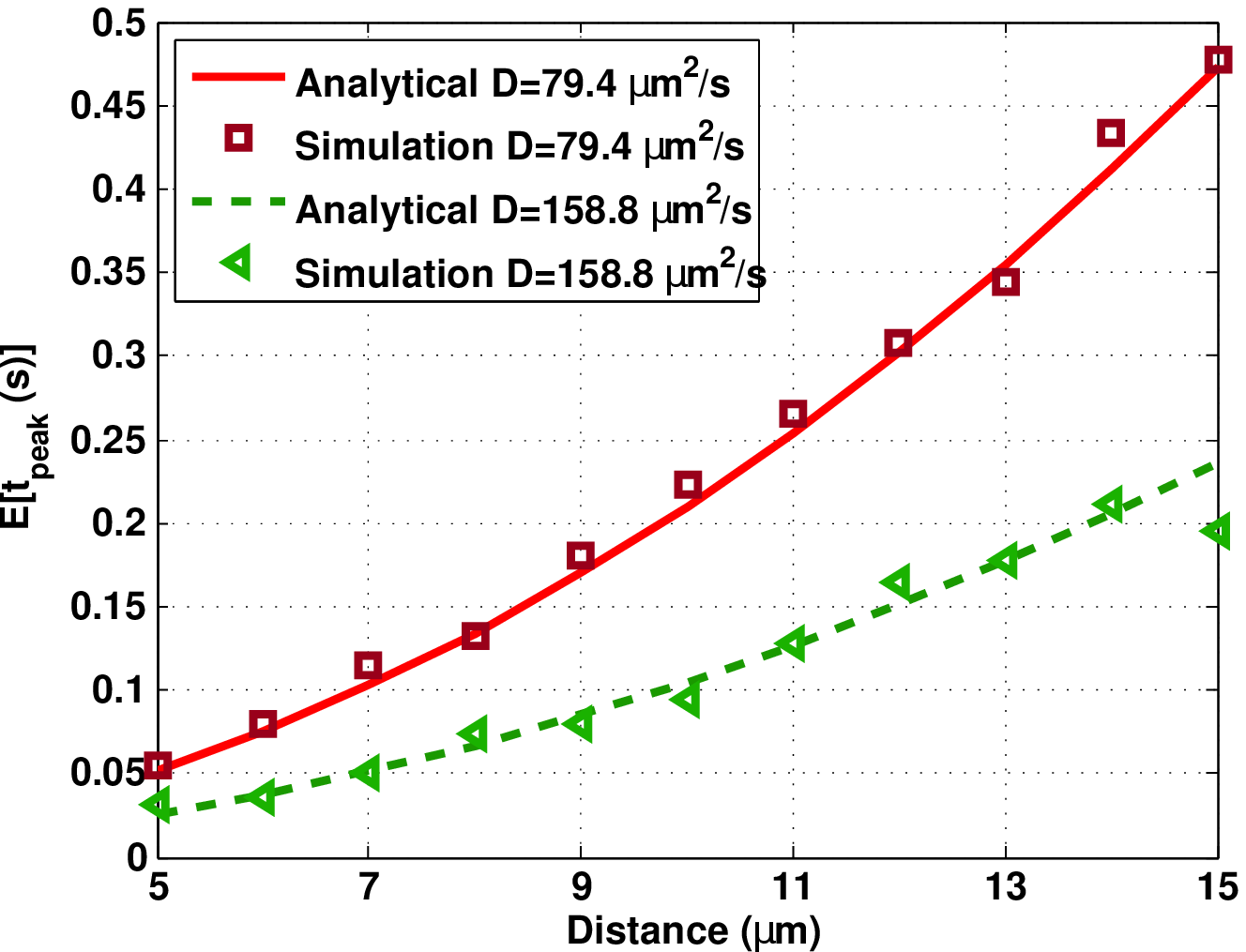}
\caption{Distance versus $\pulsept$ (${\ntxs = 5000}$ molecules, ${\rrn = 10\mu m}$).}
\label{fig:pulsePeakTime}
\end{figure}
In electromagnetic (EM) communications, $\pulsept$ is proportional to the propagation time, which is the distance divided by the wave propagation speed, hence it is proportional to $d$. In the MCvD system, due to the absorbing receiver and the diffusion dynamics, $\pulsept$ is proportional to $d^2$. Therefore, in the MCvD system, a significant factor bearing on the effectiveness of the communication is the distance parameter. 

In Fig.~\ref{fig:pulsePeakTime}, distance versus $\mathbb{E}[\pulsept]$ is presented for different $D$ values. Simulation results validate the analytical formulation. Doubling the distance makes it four times longer to observe the peak. Increasing $D$ results in a decrease in $\pulsept$; for twice $D$, we observe half of the $\pulsept$ value.

\subsection{Pulse Amplitude}
The pulse amplitude, $\pulsepn$ can be considered as the channel attenuation. If the diffusion process was evaluated without considering absorption, $\pulsepn$ would not depend on $D$, contrary to the absorbing receiver case \cite{redner2001guide, berg1993random}.  

Let us consider a pulse sent at time $t=0$, with number of emitted molecules $\ntxs$. We can obtain the maximum number of  received molecules, $\pulsepn$, by substituting $\pulsept$ into $\pdfNhit$ and multiplying it by the number of emitted molecules and $\Delta t$. 
\bea
\mathbb{E}[\pulsepn] &=& \ntxs \,\Delta t \, \pdfNhit \left|_{t=\pulsept} \right. \nonumber \\
            &=& \ntxs \,\Delta t \, \frac{\rrn}{r_0} \frac{D}{d^2} \frac{e^{-3/2}}{\sqrt{\pi/54}} \\
            &=& \ntxs \,\Delta t \, \frac{\rrn}{d+\rrn} \frac{D}{d^2} \frac{e^{-3/2}}{\sqrt{\pi/54}}.\nonumber
\label{eqn:pulsePeakValue}
\eea
The value of $\pulsepn$ in this case (considering absorption) depends on the number of emitted molecules $\ntxs$, the receiver node radius $\rrn$, distance $d$, and diffusion coefficient $D$.

For a fixed $\rrn$, we have $\pulsepn \sim 1/d^3$, and this behavior reveals the difference compared to EM communications. If we ignore fading, the amplitude of EM pulse propagating in free space decreases proportionally to the square of the transmission distance. The amplitude of a pulse in the MCvD channel, however,  decreases proportionally to the cube of the distance. We summarize the comparison of MCvD with EM in Table~\ref{tab:comparison}.

\begin{table}[tb]
\caption{Diffusion channel characteristics comparison matrix.}
\label{tab:comparison}
\centering
\begin{tabular}{llll}
\hline  Metric   & Physical Relation & Electromagnetic   & MCvD   \\
\hline  
        $\pulsept$ & Propagation Delay      & $d$     & $d^2$	 \\
        $\pulsepn$ & Path Loss      & $1/d^2$   & $1/d^3$       \\
\hline
\end{tabular}
\end{table}

\begin{figure}[!htb]
\centering
\includegraphics[width=0.9\columnwidth,keepaspectratio] {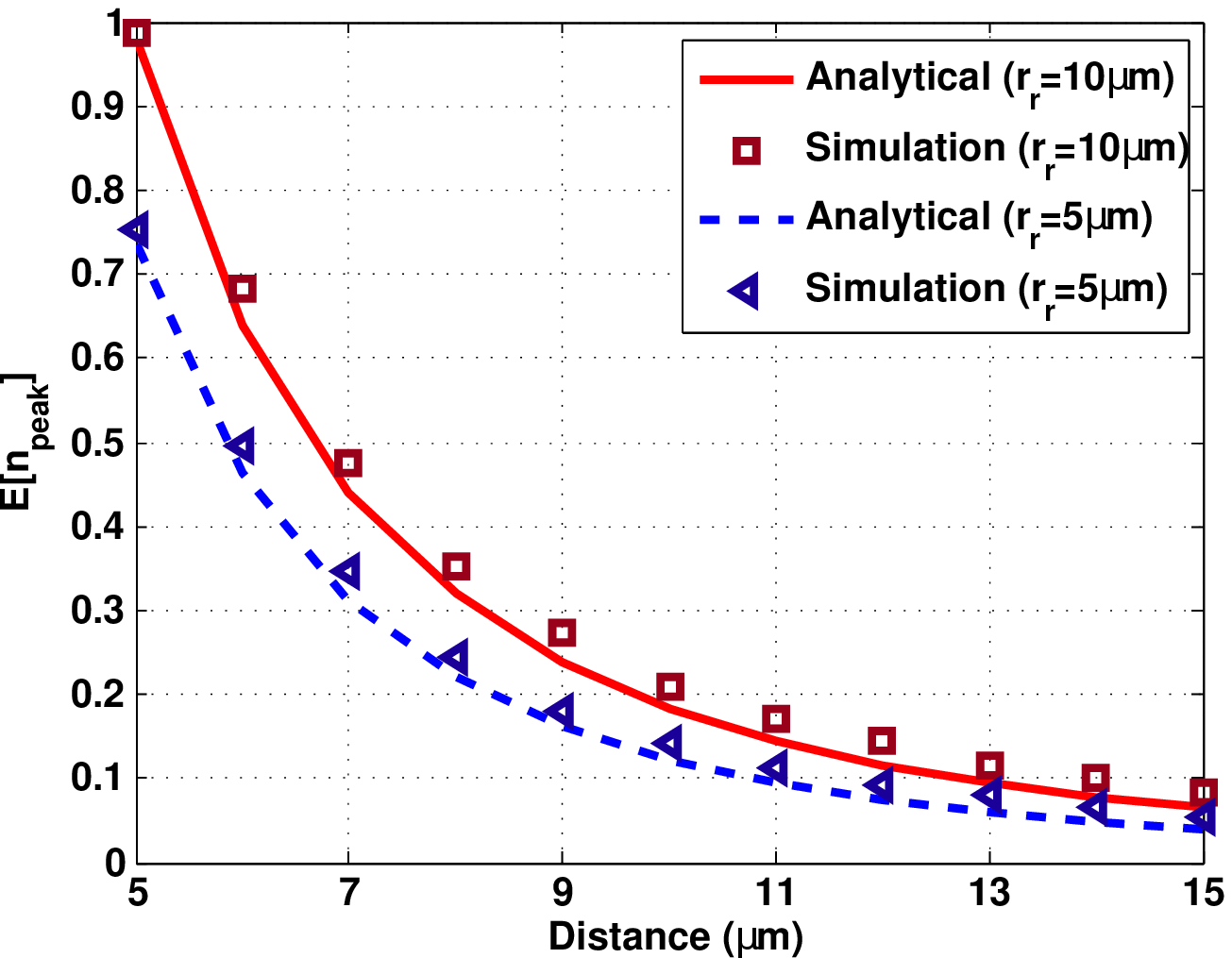}
\caption{Distance versus $\pulsepn$ (${\ntxs = 5000}$ molecules, ${D = 79.4\mu m^2/s}$).}
\label{fig:pulsePeakAmp}
\end{figure}
Fig.~\ref{fig:pulsePeakAmp} depicts distance versus pulse amplitude. We observe the close match between simulation and the analytical formulation. Increasing the distance decreases the amplitude while increasing the radius of the receiver results in higher $\mathbb{E}[\pulsepn]$. Simulation values for $\pulsepn$ are always higher than the analytical value since we take the maximum possible value in the simulations which is always expected to be larger than the mean value.

\section{Conclusion}
In this letter, we investigated the channel transfer function of a molecular communication channel for the case of an absorbing receiver. In nature, molecule absorption is commonly observed where the main means of communication is diffusion. Therefore, almost each molecule contributes to the signal once. Prior work on molecular communication, however, has not addressed this issue and has relied on concentration function. In this work, we provided analytical solutions to molecule absorption rate and fraction of absorbed molecules in a 3-D environment. Following this revelation, we presented channel metrics, pulse peak time and pulse amplitude, where simulation results verified the analytical formulations.



\bibliographystyle{IEEEtran}
\bibliography{IEEEabrv,cvdChannelCharacteristics}

\begin{thebibliography}{10}
\providecommand{\url}[1]{#1}
\csname url@samestyle\endcsname
\providecommand{\newblock}{\relax}
\providecommand{\bibinfo}[2]{#2}
\providecommand{\BIBentrySTDinterwordspacing}{\spaceskip=0pt\relax}
\providecommand{\BIBentryALTinterwordstretchfactor}{4}
\providecommand{\BIBentryALTinterwordspacing}{\spaceskip=\fontdimen2\font plus
\BIBentryALTinterwordstretchfactor\fontdimen3\font minus
  \fontdimen4\font\relax}
\providecommand{\BIBforeignlanguage}[2]{{%
\expandafter\ifx\csname l@#1\endcsname\relax
\typeout{** WARNING: IEEEtran.bst: No hyphenation pattern has been}%
\typeout{** loaded for the language `#1'. Using the pattern for}%
\typeout{** the default language instead.}%
\else
\language=\csname l@#1\endcsname
\fi
#2}}
\providecommand{\BIBdecl}{\relax}
\BIBdecl

\bibitem{tarakanov2010carbon}
A.~O. Tarakanov, L.~B. Goncharova, and Y.~A. Tarakanov, ``Carbon nanotubes
  towards medicinal biochips,'' \emph{Wiley Interdisciplinary Reviews: Nanomed.
  and Nanobiotech.}, vol.~2, no.~1, pp. 1--10, 2010.

\bibitem{akyildiz2011nanonetworks}
I.~F. Akyildiz, J.~M. Jornet, and M.~Pierobon, ``Nanonetworks: A new frontier
  in communications,'' \emph{Communications of the ACM}, vol.~54, no.~11, pp.
  84--89, 2011.

\bibitem{nakano2012molecular}
T.~Nakano, M.~J. Moore, F.~Wei, A.~V. Vasilakos, and J.~Shuai, ``Molecular
  communication and networking: Opportunities and challenges,'' \emph{IEEE
  Trans. on NanoBioscience}, vol.~11, no.~2, pp. 135--148, 2012.

\bibitem{kuran2010energy}
M.~{\c{S}}. Kuran, H.~B. Yilmaz, T.~Tugcu, and B.~{\"O}zerman, ``Energy model
  for communication via diffusion in nanonetworks,'' \emph{Nano Comm.
  Networks}, vol.~1, no.~2, pp. 86--95, 2010.

\bibitem{kim2013novelModulation}
N.-R. Kim and C.-B. Chae, ``Novel modulation techniques using isomers as
  messenger molecules for nano communication networks via diffusion,''
  \emph{IEEE Jour. Sel. Areas in Comm.}, vol.~31, no.~12, pp. 847--856, Dec.
  2013.

\bibitem{Chae_JSAC14}
N.~Farsard, N.-R. Kim, A.~Eckford, and C.-B. Chae, ``Channel and noise models
  for nonlinear molecular communication systems,'' \emph{to appear in IEEE
  Jour. Sel. Areas in Comm.}, 2014.

\bibitem{tyrrell1984diffusion}
H.~J.~V. Tyrrell and K.~Harris, \emph{Diffusion in liquids, A theoretical and
  experimental study}.\hskip 1em plus 0.5em minus 0.4em\relax Butterworth
  Publishers, Stoneham, MA, 1984.

\bibitem{redner2001guide}
S.~Redner, \emph{A guide to first-passage processes}.\hskip 1em plus 0.5em
  minus 0.4em\relax Cambridge University Press, 2001.

\bibitem{schulten2000lectures}
K.~Schulten and I.~Kosztin, ``Lectures in theoretical biophysics,''
  \emph{University of Illinois}, vol. 117, 2000.

\bibitem{berg1993random}
H.~C. Berg, \emph{Random walks in biology}.\hskip 1em plus 0.5em minus
  0.4em\relax Princeton University Press, 1993.

\end{thebibliography}
%

%



%




\end{document}